\documentclass[twocolumn,superscriptaddress,amsmath,amssymb,aps,prl]{revtex4-1}

\usepackage{graphicx}
\usepackage{dcolumn}
\usepackage{bm}
\newcommand{\tabincell}[2]{\begin{tabular}{@{}#1@{}}#2\end{tabular}}

\usepackage[colorlinks,urlcolor=blue,citecolor=blue,linkcolor=blue]{hyperref}

\begin{document}
	\title{Complex contact interaction for systems with short-range two-body loss}

	\author{Ce Wang}
	\affiliation{Institute for Advanced Study, Tsinghua University, Beijing 100084, China}
	\author{Chang Liu}
	\affiliation{Institute for Advanced Study, Tsinghua University, Beijing 100084, China}
	\author{Zhe-Yu Shi}
	\email{zyshi@lps.ecnu.edu.cn}
	\affiliation{State Key Laboratory of Precision Spectroscopy, East China Normal University, Shanghai 200062, China}
	\date{\today}
	
	\begin{abstract}
	Contact interaction is a fundamental concept that appears in various areas of physics. It simplifies physical models by replacing the detailed short-range interaction with a zero-range contact potential which reproduces the same low-energy scattering parameter, {\it i.e.} the $s$-wave scattering length. In this work, we generalize this concept to a system with short-range two-body loss. We show that the short-range two-body loss can effectively be replaced by a zero-range {\it complex} contact potential with proper regularization characterized by a {\it complex} scattering length. We develop appropriate ways to regularize this potential in the Lindblad master equation and apply them to the dynamic problem of Bose-Einstein condensate with weak interaction and two-body loss.
	\end{abstract}

	\maketitle
	
	Separation of scales appears in many physical systems. It allows us to construct simple models that are able to capture the most fundamental picture of the physics effectively. For example, separation of length scales happens in systems such as ultracold atomic gases and nuclear systems where the ranges of the interparticle interactions are much smaller than other length scales such as the interparticle distances and the thermal de Broglie wavelength. These complicated short-range interactions can then be replaced by a zero-range contact potential, once the contact potential reproduces the same physical behavior for a low-energy collision process.

	In scattering theory, the low-energy scattering data is described by a single $s$-wave scattering length $a$~\cite{landau2013quantum}. Given the $s$-wave scattering length, there are three major approaches that can describe the zero-range contact interaction in the literature, which include Bethe-Peierls model, pseudopotential, and renormalized delta-potential. We briefly introduce them in the following.

	{\it Behte-Peierls model.-} In their study of the scattering theory of deuterons, Bethe and Peierls suggest that the effect of a short-range potential $V(r)$ may be replaced by a boundary condition at $r=0$~\cite{bethe1935quantum}. It is shown that the zero-energy solution for the two-body relative wavefunction is $\varphi(\mathbf{r}_\text{rel})=r_\text{rel}^{-1}-a^{-1}$ outside the interaction range $r_0$. Thus if we are only interested in the low-energy physics in such systems, the interaction can be replaced by a boundary condition on the many-body wavefunction~\cite{bosons},
	\begin{align}
		\psi(\underline{\mathbf{r}}_N)\simeq\left(\frac{1}{r_{ij}}-\frac{1}{a}\right)A(\underline{\mathbf{r}}_N^{(ij)},\mathbf{R}_{ij}),\quad r_{ij}\rightarrow0,\label{real_bp}
	\end{align}
	where $A$ could be an arbitrary function, $\mathbf{R}_{ij}=\frac{\mathbf{r}_i+\mathbf{r}_j}{2}$ and $\mathbf{r}_{ij}=\mathbf{r}_i-\mathbf{r}_j$ are the center of mass and relative coordinates of particle $i$ and $j$, $\underline{\mathbf{r}}_N$ represents all the coordinates in $\{\mathbf{r}_1,\ldots,\mathbf{r}_N\}$, $\underline{\mathbf{r}}_N^{(ij)}$ represents all the coordinates except $\mathbf{r}_i$ and $\mathbf{r}_j$.

	{\it Pseudopotential.-} First introduced by Fermi, the pseudopotential models the short-range interaction through a delta-potential and an extra operator which regularizes the wavefunction near the origin~\cite{fermi1936motion,breit1947scattering,blatt1991theoretical},
	\begin{align}
		U(\mathbf{r})=\frac{4\pi\hbar^2 a}{m}\delta({\mathbf{r}})\partial_rr,\label{real_pseudo}
	\end{align}
	with $m$ the particle mass. It can be shown that this pseudopotential is equivalent to posing the boundary condition~\eqref{real_bp} at the origin~\cite{blatt1991theoretical,huang1957quantum}.

	{\it Renormalized delta-potential.-} Another way to regularize the delta-potential is to use the renormalization method developed in quantum field theory. Given $V(\mathbf{r})=g\delta(\mathbf{r})$, one can calculate the on shell two-body T-matrix $t(E)$ and compare it with the low-energy scattering amplitude $f(E)$ via $t(E)=-\frac{4\pi\hbar^2}{m}f(E)$. This relates the coupling constant $g$ to the $s$-wave scattering length $a$ through renormalization relation~\cite{randeria1995bose},
	\begin{align}
		\frac{1}{g}=\frac{m}{4\pi\hbar^2a}-\frac{1}{\Omega}\sum_{\mathbf{k}}\frac{1}{2\epsilon_\mathbf{k}}.
	\end{align}
	Here $\epsilon_\mathbf{k}=\frac{\hbar^2k^2}{2m}$ is the single particle dispersion and $\Omega$ is the system volume. It is worth noting that the momentum summation in the R.H.S. will leads to a ultraviolet divergence than needs to be properly cancelled in any practical calculation.

	These equivalent descriptions are the foundation of many successful theories, from the ground state energy correction of weakly interacting Bose-Einstein condensates (BECs)~\cite{lee1957many,lee1957eigenvalues} to the BEC-BCS crossover in two-component Fermi gases~\cite{eagles1969possible,leggett1980diatomic,randeria1995bose}. Furthermore, in ultracold atomic gases, the Feshbach resonance technique~\cite{inouye1998observation,courteille1998observation,chin2010feshbach} provides a tool for controlling the interparticle potential between atoms by varying the their scattering length, which allows the the study of various many-body effects on quantum systems.

	In this work, we generalize the concept of contact potential to an open system with short-range two-body loss where the particle collision process becomes inelastic. We discuss the structure of the general Lindblad master equation for systems with finite-range interactions and two-body losses. By taking the limit of interaction and loss range $r_0\rightarrow0$, we show that the only important low-energy parameter remains is a complex scattering length $a_c$. We further develop three equivalent methods to regularize or renormalize the contact (zero-range) interactions and two-body losses in the Lindblad master equation, which are listed in table~\ref{table1}. We then apply our model to calculate the dynamics of a BEC with weak interaction and two-body loss. The experimental methods of tuning the complex scattering length $a_c$ is also discussed~\cite{sm}.

	{\it The Lindblad master equation.-} Consider an open system of interacting bosons subject to (finite-range) two-body losses, the evolution of the density matrix $\hat{\rho}$ is governed by the Lindblad master equation $\partial_t\hat{\rho}=\mathcal{L}\hat{\rho}$ with the Lindbladian ($\hbar=1$)~\cite{breuer2002theory}
	\begin{align}
		\mathcal{L}\hat{\rho}=\frac{1}{i}[\hat{H},\hat{\rho}]-\frac{1}{2}\int_{\mathbf{r}_1,\mathbf{r}_2}\!\!\!\!\!V_i(r_{12})\{\hat{\psi}^\dagger_{\mathbf{r}_1}\hat{\psi}^\dagger_{\mathbf{r}_2}\hat{\psi}_{\mathbf{r}_2}\hat{\psi}_{\mathbf{r}_1},\hat{\rho}\}+\mathcal{J}\hat{\rho},\nonumber
	\end{align}
	where $\hat{\psi}_{\mathbf{r}}$ is the annihilation operator at position $\mathbf{r}$, $\hat{H}$ is the usual Hermitian Hamiltonian of interacting bosons,
	\begin{align}
		\hat{H}=-\int_\mathbf{r}\hat{\psi}^\dagger_\mathbf{r}\frac{\nabla^2}{2m}\hat{\psi}_\mathbf{r}+\frac{1}{2}\int_{\mathbf{r}_1,\mathbf{r}_2}\!\!\!\!\!V_r(r_{12})\hat{\psi}^\dagger_{\mathbf{r}_1}\hat{\psi}^\dagger_{\mathbf{r}_2}\hat{\psi}_{\mathbf{r}_2}\hat{\psi}_{\mathbf{r}_1}.\label{real_hamiltonian}
	\end{align}
	We assume the two-body loss rate $V_i$ is a function that depends on the interparticle distance.  The recycling term $\mathcal{J}\hat{\rho}$ is then given by
	\begin{align}
		\mathcal{J}\hat{\rho}=\int_{\mathbf{r}_1,\mathbf{r}_2}\!\!\!\!\!V_i(r_{12})\hat{\psi}_{\mathbf{r}_1}\hat{\psi}_{\mathbf{r}_2}\hat{\rho}\hat{\psi}^\dagger_{\mathbf{r}_2}\hat{\psi}^\dagger_{\mathbf{r}_1}.
	\end{align}
	The interaction $V_r$ and the two-body loss rate $V_i$ are assume to be finite ranged and vanish at $r>r_0$. It is also required that $V_i\geq0$ inside $r_0$, which is necessary to guarantee the positive definiteness of the density matrix. 

	The master equation may be regarded as the evolution under a non-Hermitian Hamiltonian $\hat{H}_\text{eff}$ together with the recycling term, {\it i.e.} $\partial_t\hat{\rho}=\frac{1}{i}(\hat{H}_\text{eff}\hat{\rho}-\hat{\rho}\hat{H}_\text{eff}^\dagger)+\mathcal{J}\hat{\rho}$, where the $\hat{H}_\text{eff}$ is similar to the Hermitian Hamiltonian $\hat{H}$ but with the real potential $V_r$ replaced by a complex one $V_c=V_r-iV_i$.
	
	{\it The complex scattering length.-} The special form of the jump operator $\hat{\psi}_{\mathbf{r}_1}\hat{\psi}_{\mathbf{r}_2}$ leads to a hierarchical structure of the Lindbladian $\mathcal{L}$. To see this, note that the bosonic Fock space naturally defines orthogonal projections $\hat{P}_l,\ l=0,1,2\ldots$ which project any state to the $l$-boson subspace $\mathcal{H}_l$. For any linear operator $\hat{O}$, we thus have decomposition $\hat{O}=\sum_{j,l}\hat{O}_{jl}$ with $\hat{O}_{jl}\equiv\hat{P}_j\hat{O}\hat{P}_l$ an operator that maps a state in $\mathcal{H}_l$ to $\mathcal{H}_j$. Because the jump operator $\hat{\psi}_{\mathbf{r}_1}\hat{\psi}_{\mathbf{r}_2}$ always annihilates two particles, one can show that the master equation may be decomposed to a series of hierarchy equations for $\hat{\rho}_{jl}$,
	\begin{align}
		\partial_t\hat{\rho}_{jl}=\frac{1}{i}(\hat{H}_\text{eff}\hat{\rho}_{jl}-\hat{\rho}_{jl}\hat{H}_\text{eff})+\mathcal{J}\hat{\rho}_{j+2,l+2}.\label{hierarchy}
	\end{align}

	The hierarchical structure allows us to consider a ``two-body'' problem in the presence of two-body loss. If we start with an initial density matrix $\hat{\rho}(0)$ that contains two bosons, {\it i.e.} $\hat{\rho}(0)=\hat{\rho}_{22}(0)$. It is clear from eq.~\eqref{hierarchy} that the only nonvanishing blocks of $\hat{\rho}(t)$ will be $\hat{\rho}_{22}$ and $\hat{\rho}_0$, which satisfy,
	\begin{align}
		\partial_t\hat{\rho}_{22}&=\frac{1}{i}(\hat{H}_\text{eff}\hat{\rho}_{22}-\hat{\rho}_{22}\hat{H}_\text{eff}^\dagger),\\
		\partial_t\hat{\rho}_{00}&=\mathcal{J}\hat{\rho}_{22}=\partial_t\text{tr}\hat{\rho}_{22}.
	\end{align}

	We see that the evolution of the two-particle density matrix $\hat{\rho}_{22}$ is fully described by the non-Hermitian Hamiltonian $\hat{H}_\text{eff}$. This means the ``two-body'' problem may be solved in the same manner as the usual two-body problem except that the potential $V_c(r)$ is complex. Consider the $s$-wave zero-energy wavefunction in relative coordinates $\varphi(r)$. It is then clear that
	\begin{align}
	 	\varphi(r)=\frac{1}{r}-\frac{1}{a_c},\text{ for }r\geq r_0,\label{complex_a}
	\end{align} 
	because the system is non-interacting in this region. 

	Eq.~\eqref{complex_a} gives the definition of the complex scattering length $a_c$. Furthermore, it can be shown that $\text{Im}(a_c^{-1})=m\int_0^{r_0}r^2drV_i(r)|\varphi(r)|^2$~\cite{ai}. Together with the constrain $V_i\geq0$, we conclude that $\text{Im}(a_c)$ is always negative in the presence of two-body loss. We thus write $a_c$ as $a_c=a_r+ia_i$ with $a_i<0$.

	\begin{table*}[htbp]
	\centering  
	\resizebox{2\columnwidth}{!}{
	\begin{tabular}{|c|c|c|c|}  
		\hline  
		& & & \\[-6pt]  
		&contact interaction&complex contact interaction&recycling term \\  

		\hline
		& & & \\[-6pt]   

		\tabincell{c}{Bethe-Peierls\\model}&
		\tabincell{c}{$\psi(\underline{\mathbf{r}}_N)\simeq(\frac{1}{r_{\alpha\beta}}-\frac{1}{a})A(\mathbf{\underline{r}}_N^{(\alpha\beta)},\mathbf{R}_{\alpha\beta})$}&
		\tabincell{c}{$\rho_{jl}(\mathbf{\underline{r}}_j,\mathbf{\underline{r}}_l')\simeq(\frac{1}{r_{\alpha\beta}}-\frac{1}{a_c})(\frac{1}{r_{\mu\nu}}-\frac{1}{a_c^*})$\\  $\times B_{jl}(\mathbf{\underline{r}}_j^{(\alpha\beta)},\mathbf{R}_{\alpha\beta};\mathbf{\underline{r}}_l'{}^{(\mu\nu)},\mathbf{R}_{\mu\nu}')$}&
		\tabincell{c}{$\displaystyle\text{Im}\left(\frac{4\pi\hbar^2}{ma_c}\right)\sqrt{(j+2)(j+1)(l+2)(l+1)}$\\$\displaystyle\times\int_\mathbf{R} B_{j+2,l+2}(\mathbf{\underline{r}}_j,\mathbf{R};\mathbf{\underline{r}}_l',\mathbf{R})$}\\

		& & & \\[-6pt] 
		\hline
		& & & \\[-6pt]   

		\tabincell{c}{pseudopotential}&
		$\displaystyle U(\mathbf{r})=\frac{4\pi \hbar^2 a}{m}\delta(\mathbf{r}) \partial_{r}r$&
		$\displaystyle U_c(\mathbf{r})=\frac{4\pi \hbar^2 a_c}{m}\delta(\mathbf{r}) \partial_{r}r$&
		\tabincell{c}{$\displaystyle\frac{4\pi\hbar^2|a_i|}{m}\sqrt{(j+2)(j+1)(l+2)(l+1)}$\\
		$\displaystyle\times\int_{\mathbf{R},\mathbf{r},\mathbf{r}'}\!\!\!\!\!\delta(\mathbf{r})\delta(\mathbf{r}')\partial_r r\partial_{r'}r'\rho_{j+2,l+2}$}\\

		& & & \\[-6pt]	
		\hline 
		& & & \\[-6pt]

	   \tabincell{c}{renormalization\\relation}&
	   $\displaystyle\frac{m}{4\pi\hbar^2 a}=\frac{1}{g} + \frac{1}{\Omega} \sum_{\mathbf{k}}\frac{1}{2\epsilon_\mathbf{k}}$&
	   $\displaystyle\frac{m}{4\pi\hbar^2 a_c}  = \frac{1}{g-i\gamma} + \frac{1}{\Omega} \sum_{\mathbf{k}}\frac{1}{2\epsilon_\mathbf{k}}$&
	   $\displaystyle \gamma\int_\mathbf{r} \hat{\psi}^2_\mathbf{r}\hat{\rho}\hat{\psi}^{\dagger 2}_\mathbf{r}$ \\

		\hline
		
	\end{tabular}
	}
	\caption{Three approaches regularizing the contact interaction and their complex analogs for zero-range two-body loss. We denote coordinates $\{\mathbf{r}_1,\mathbf{r}_2\ldots,\mathbf{r}_N\}$ by $\mathbf{\underline{r}}_N$. $\mathbf{\underline{r}}_N^{(\alpha\beta)}$ stands for all the coordinates in $\underline{\mathbf{r}}_N$ except the two with indices $\alpha,\beta$. $\mathbf{R}_{\alpha\beta}\equiv(\mathbf{r}_i+\mathbf{r}_j)/2$ and $\mathbf{r}_{\alpha\beta}\equiv\mathbf{r}_\alpha-\mathbf{r}_\beta$ stand for the center of mass and relative coordinates of particles $\alpha$ and $\beta$ respectively. The density matrix $\rho_{j+2,l+2}$ in the middle right cell stands for $\rho_{j+2,l+2}(\mathbf{\underline{r}}_j,\mathbf{R}+\mathbf{r}/2,\mathbf{R}-\mathbf{r}/2;\mathbf{\underline{r}}_l',\mathbf{R}+\mathbf{r}'/2,\mathbf{R}-\mathbf{r}'/2)$.}  
	\label{table1}  
	\end{table*}

	{\it Complex Bethe-Peierls model.-} To generalize the Bethe-Peierls boundary condition, we first write the Lindblad equation in the first quantization formalism. Acting $\langle \underline{\mathbf{r}}_j|\cdot|\underline{\mathbf{r}}'_l\rangle$ on both sides of eq.~\eqref{hierarchy} ($|\underline{\mathbf{r}}_l\rangle\equiv\frac{1}{\sqrt{l!}}\hat{\psi}_{\mathbf{r}_1}^\dagger\ldots\hat{\psi}_{\mathbf{r}_l}^\dagger|0\rangle$), we obtain
	\begin{align}
		\partial_t\rho_{jl}=\frac{1}{i}\left(H_\text{eff}(\underline{\mathbf{r}}_j)-H_\text{eff}^\dagger(\underline{\mathbf{r}}_l')\right)\rho_{jl}+\mathcal{J}\rho_{j+2,l+2},\label{lindblad_1st}
	\end{align}
	where $\rho_{jl}(\underline{\mathbf{r}}_j,\underline{\mathbf{r}}_l')\equiv\langle \underline{\mathbf{r}}_j|\hat{\rho}_{jl}|\underline{\mathbf{r}}_l'\rangle$ is the first quantized density matrix, $H_\text{eff}(\underline{\mathbf{r}}_j)=\sum_{\alpha=1}^j-\frac{\nabla_\alpha^2}{m}+\sum_{1\leq\alpha<\beta\leq j}V_c(r_{\alpha\beta})$ is the first quantized Hamiltonian. The recycling term is given by
	\begin{align}
		&\mathcal{J}\rho_{j+2,l+2}=\sqrt{(j+2)(j+1)(l+2)(l+1)}\nonumber\\
		&\qquad\quad\times\int_{\mathbf{x},\mathbf{y}}\!\!\!\!\!V_i(|\mathbf{x}-\mathbf{y}|)\rho_{j+2,l+2}(\underline{\mathbf{r}}_j,\mathbf{x},\mathbf{y};\underline{\mathbf{r}}_l',\mathbf{x},\mathbf{y}).\label{recycling_1st}
	\end{align}

	From eq.~\eqref{lindblad_1st}, we notice that in the region where all the particles are apart from each other such that $r_{\alpha\beta},r'_{\alpha\beta}>r_0$ for all possible distinct pairs $\alpha$, $\beta$, the evolution of $\rho_{jl}$ is governed by a noninteracting $H_\text{eff}$ plus the recycling term $\mathcal{J}\rho_{j+2,l+2}$. In the zero-range limit $r_0\rightarrow0$, this region fills the whole domain of $\rho_{jl}$, one thus expects that the effect of the complex interaction $V_c$ can be replaced by a boundary condition at $r_{\alpha\beta}\rightarrow0$.

	To be more concrete, we consider a system with mean inter-particle distance $d$ and energy per particle $\frac{k^2}{2m}$, and focus on the density matrix with a pairs of particles ($\alpha$ and $\beta$) close to each other such that $r_{\alpha\beta}\ll d,k^{-1}$. In this region, the two-body scattering process dominates and every other terms in eq.~\eqref{lindblad_1st} besides the two-body relative kinetic energy and interaction $V_c(r_{\alpha\beta})$ can be ignored~\cite{drop}. Then Lindblad equation then reduces to
	\begin{align}
		0\simeq-\frac{\nabla^2_{\mathbf{r}_{\alpha\beta}}}{2m}\rho_{jl}+V_c(r_{\alpha\beta})\rho_{jl},
	\end{align}
	which is nothing but the zero-energy two-body Schr\"{o}dinger equation in the relative coordinate $\mathbf{r}_{\alpha\beta}$.

	Because of the centrifugal barrier of higher partial waves, $\rho_{jl}$ is dominated by the $s$-wave two-body wave function $\varphi(r)$. We thus have $\rho_{jl}\propto\varphi(r_{\alpha\beta})$ when $r_{\alpha\beta}\rightarrow0$. The same proof may also be applied to the region $r'_{\mu\nu}\ll d,k^{-1}$, which leads to following asymptotic form of $\rho_{jl}(\underline{\mathbf{r}}_j,\underline{\mathbf{r}}_l)$ when $r_{\alpha\beta},r'_{\mu\nu}\rightarrow0$,
	\begin{align}
		\rho_{jl}\simeq\varphi(r_{\alpha\beta})\varphi(r_{\mu\nu})B_{jl}(\underline{\mathbf{r}}_j^{(\alpha\beta)},\mathbf{R}_{\alpha\beta};\underline{\mathbf{r}}'_l{}^{(\mu\nu)},\mathbf{R}'_{\mu\nu})\label{asy}
	\end{align}
	with $B_{jl}$ an arbitrary function.

	Taking the limit of $r_0\rightarrow0$, we obtain the boundary condition,
	\begin{align}
		\rho_{jl}&\simeq\left(\frac{1}{r_{\alpha\beta}}-\frac{1}{a_c}\right)\left(\frac{1}{r_{\mu\nu}}-\frac{1}{a_c^*}\right)\label{complex_bp}\\
		&\times B_{jl}(\underline{\mathbf{r}}_j^{(\alpha\beta)},\mathbf{R}_{\alpha\beta};\underline{\mathbf{r}}'_l{}^{(\mu\nu)},\mathbf{R}'_{\mu\nu}),\quad r_{\alpha\beta},r'_{\mu\nu}\rightarrow0\nonumber.
	\end{align}

	The recycling term can be calculated by substituting eq.~\eqref{asy} into eq.~\eqref{recycling_1st}, which leads to
	\begin{align}
		\mathcal{J}\rho_{j+2,l+2}&=\text{Im}\left(\frac{4\pi\hbar^2}{ma_c}\right)\sqrt{(j+2)(j+1)(l+2)(l+1)}\nonumber\\
		&\times\int_{\mathbf{R}}B_{j+2,l+2}(\underline{\mathbf{r}}_j,\mathbf{R};\underline{\mathbf{r}}_l',\mathbf{R})\label{rec_bp}
	\end{align}
	where we restored $\hbar$.

	The boundary condition~\eqref{complex_bp} together with the recycling term~\eqref{rec_bp} determine the evolution of density matrix $\rho_{jl}$ in the zero-range limit. They thus can be viewed as the complex analog of the Bethe-Peierls boundary condition~\eqref{real_bp}.

	{\it Complex pseudopotential.-} Given the boundary condition~\eqref{complex_bp}, it is straightforward to apply the standard regularization method~\cite{blatt1991theoretical,huang1957quantum} and show that the short-range complex interaction $V_c$ ($V_c^*$) in $H_\text{eff}$ ($H_\text{eff}^\dagger$) can also be replaced by a complex pseudopotential $U_c$ ($U_c^*$) with
	\begin{align}
		U_{c}(\mathbf{r})=\frac{4\pi\hbar^2a_c}{m}\delta(\mathbf{r})\partial_rr.\label{complex_pseudo}
	\end{align}

	Similarly, the recycling term~\eqref{rec_bp} can be written in terms of the reguarlized operators
	\begin{align}
		\mathcal{J}\rho_{j+2,l+2}&=\frac{4\pi\hbar^2|a_i|}{m}\sqrt{(j+2)(j+1)(l+2)(l+1)}\nonumber\\
		&\times\int_{\mathbf{R},\mathbf{r},\mathbf{r}'}\!\!\!\!\!\delta(\mathbf{r})\delta(\mathbf{r}')\partial_rr\partial_{r'}r'\rho_{j+2,l+2},\label{rec_pseudo}
	\end{align}
	where $\hat{\rho}_{j+2,l+2}$ stands for $\rho_{j+2,l+2}(\underline{\mathbf{r}}_j,\mathbf{R}+\frac{\mathbf{r}}{2},\mathbf{R}-\frac{\mathbf{r}}{2};\underline{\mathbf{r}}_l',\mathbf{R}+\frac{\mathbf{r}'}{2},\mathbf{R}-\frac{\mathbf{r}'}{2})$.


	{\it Renormalized contact potential.-} Following the conventional renormalization approach, we first write the short-range complex potential $V_c$ as a delta-potential,
	\begin{align}
		V_c=(g-i\gamma)\delta(\mathbf{r}),\label{complex_delta}
	\end{align}
	with $g$ ($\gamma$) being the real (imaginary) coupling constant.

	It is then straightforward to calculate the two-body scattering amplitude~\cite{book_zhai},
	\begin{align}
		f(k)=-\frac{m}{4\pi}\frac{1}{(g-i\gamma)^{-1}+\frac{1}{\Omega}\sum_{\mathbf{k}}(2\epsilon_\mathbf{k})^{-1}-\frac{ikm}{4\pi}}.
	\end{align}
	Compare this formula with the standard low-energy expansion of the scattering amplitude $f(k)=-1/(a_c^{-1}+ik)$, we find renormalization relation,
	\begin{align}
		\frac{1}{g-i\gamma}=\frac{1}{g_0-i\gamma_0}-\frac{1}{\Omega}\sum_\mathbf{k}\frac{1}{2\epsilon_\mathbf{k}},\label{complex_renormalization}
	\end{align}
	where we have defined $g_0-i\gamma_0\equiv\frac{4\pi\hbar^2 a_c}{m}$ being the renormalized complex coupling constant. And the second quantized recycling term is simply
	\begin{align}
		\mathcal{J}\hat{\rho}=\gamma\int_\mathbf{r}\hat{\psi}^2_\mathbf{r}\hat{\rho}\hat{\psi}^{\dagger2}_\mathbf{r}.~\label{complex_recycling}
	\end{align}


	We list the results for the three regularization approaches in table~\ref{table1}. It is worth noting that the renormalization relation~\eqref{complex_renormalization} has already been used for the calculation of non-Hermitian models in two recent works~\cite{iskin2021non,zhou2021effective}. However, both works focus on the calculation of non-Hermitian Hamiltonian $H_\text{eff}$, and neither the correct form (eq.~\eqref{complex_recycling}) nor the effect of the recycling term $\mathcal{J}\hat{\rho}$ are addressed.

	{\it Application to Bose gases.-} To demonstrate the validity of our regularized model, we study the quench dynamics of BECs subjected to weak interaction and loss, {\it i.e.} $n|a_c|^3\ll1$ where $n$ is the boson density.

	We shall use the renormalized delta potential approach for this many-body problem. Write the original Lindbladian in momentum space, we obtain
	\begin{align}
		H_\text{eff}=\sum_\mathbf{k}\epsilon_\mathbf{k}\hat{a}_\mathbf{k}^\dagger \hat{a}_\mathbf{k}+\frac{g-i\gamma}{2\Omega}\sum_{\mathbf{k},\mathbf{k}',\mathbf{p}}\hat{a}^\dagger_{\mathbf{k+p}}\hat{a}^\dagger_{\mathbf{k}'-\mathbf{p}}\hat{a}_{\mathbf{k}'}\hat{a}_\mathbf{k},
	\end{align}
	and the recycling term
	\begin{align}
		\mathcal{J}\hat{\rho}=\frac{\gamma}{\Omega}\sum_{\mathbf{k},\mathbf{k}',\mathbf{p}}\hat{a}_{\mathbf{k}'}\hat{a}_\mathbf{k}\hat{\rho}\hat{a}^\dagger_{\mathbf{k'-p}}\hat{a}^\dagger_{\mathbf{k}+\mathbf{p}},
	\end{align}
	where $a_\mathbf{k}^\dagger\equiv\frac{1}{\sqrt{\Omega}}\int_\mathbf{r}e^{i\mathbf{k}\cdot\mathbf{r}}\psi_\mathbf{r}^\dagger$.

	We consider a system of $N$ bosons initially condense in the zero momentum state, such that a large fraction of bosons still remains in the condensate when $t$ is small, {\it i.e.} the depletion $(N-N_0)/N\ll1$ where $N_0\equiv\langle \hat{a}_0^\dagger \hat{a}_0\rangle$ is the number of particles in the condensate. Then we may apply the Bogoliubov approximation~\cite{bogoliubov1947theory} and substitute $\hat{a}_0,\hat{a}_0^\dagger$ in the Lindbladian by $\sqrt{N-\sum_{\mathbf{k}\neq0}\hat{a}_\mathbf{k}^\dagger\hat{a}_\mathbf{k}}$. This leads to a quadratic Bogoliubov Lindbladian that describes the dynamics of the non-condensed bosons,
	\begin{align}
		\mathcal{L}_B\hat{\rho}'=&\frac{1}{i}\left[\hat{H}_B,\hat{\rho}'\right]-2\gamma_0 n\sum_{\mathbf{k}\neq0}\left\{\hat{a}_\mathbf{k}^\dagger \hat{a}_\mathbf{k},\hat{\rho}'\right\}\nonumber\\
		&+4\gamma_0 n\sum_{\mathbf{k}\neq0}\hat{a}_\mathbf{k}\hat{\rho}'\hat{a}_\mathbf{k}.
	\end{align}
	Here $\hat{\rho}'$ is the reduced density matrix for the non-condensed bosons. We see that the Lindbladian $\mathcal{L}_B$ describes an open system governed by $\hat{H}_B$ and single particle loss with loss rate $4\gamma_0n$. Here $H_B$ is a Hermitian Hamiltonian,
	\begin{align}
		\hat{H}_B=\sum_{\mathbf{k}\neq0}\left((\epsilon+g_0n)\hat{a}_\mathbf{k}^\dagger\hat{a}_\mathbf{k}+\frac{g_0n-i\gamma_0 n}{2}\hat{a}_\mathbf{k}^\dagger \hat{a}_{-\mathbf{k}}^\dagger+h.c.\right).
	\end{align}
	Similar to the conventional Bogoliubov approximation approach~\cite{pethick2008bose}, we replaced all the bare coupling constants $g,\gamma$ by renormalized values $g_0,\gamma_0$.

	{\it Remarks on $\mathcal{L}_B$.-}  We emphasize that the recycling term $\mathcal{J}\hat{\rho}$ is essential for deriving the correct many-body Lindbladian $\mathcal{L}_B$, as part of the recycling term such as $\frac{\gamma}{\Omega}\hat{a}_0\hat{a}_0\hat{\rho}\hat{a}^\dagger_\mathbf{k}\hat{a}^\dagger_\mathbf{-k}$ becomes $\gamma n\hat{\rho}\hat{a}^\dagger_\mathbf{k}\hat{a}^\dagger_\mathbf{-k}$ and constitutes the Hermitian Hamiltonian $H_B$ after the approximation. It shows that the recycling term $\mathcal{J}\hat{\rho}$ indeed plays an important role in the many-body dynamics and it is crucial to regularize it accordingly. Moreover, we note that the total density $n$ is time-dependent due to the breakdown of particle number conservation. However, for systems with $n|a_c|^3\ll1$, one can simply substitute it by the mean-field value $n(t)=n(0)/(1+\gamma_0n(0)t)$, which gives the correct results to the order we desire (see the derivation below).

	Equipped with these remarks, the quadratic Lindbladian $\mathcal{L}_B$ is easy to solve. For example, we may consider the dynamics of the SU(1,1) generators for the conventional Bogoliubov Hamiltonian, $A_0^\mathbf{k}=\frac{1}{2}(N_\mathbf{k}+N_{-\mathbf{k}}+1)$, $A_1^\mathbf{k}=\frac{1}{2}(\hat{a}_\mathbf{k}^\dagger\hat{a}_{\mathbf{-k}}^\dagger+h.c.)$, and $A_2^\mathbf{k}=\frac{1}{2i}(\hat{a}_\mathbf{k}^\dagger\hat{a}_{\mathbf{-k}}^\dagger-h.c.)$~\cite{chen2020many}. 

	The dynamics of $A^\mathbf{k}_i$ may be calculated by $\frac{d}{dt}\langle A_i^\mathbf{k}\rangle=\text{tr}(\partial_t\hat{\rho}'A_i^\mathbf{k})=\text{tr}(\mathcal{L}_B\hat{\rho}' A_i^\mathbf{k})$, which leads to a closed matrix equation,
	\begin{align}
		\dot{\mathbf{A}}^\mathbf{k}=-2\left(\begin{array}{ccc}
		2\gamma_0 n&\gamma_0 n&-g_0n\\
		\gamma_0 n&2\gamma_0 n&\epsilon_\mathbf{k}+g_0n\\
		-g_0n&-\epsilon_\mathbf{k}-g_0n&2\gamma_0 n
	\end{array}
	\right)\mathbf{A}^\mathbf{k}+\left(\begin{array}{c}
		2\gamma_0 n\\
		0\\
		0
	\end{array}\right)\label{matrix_eq}
	\end{align}
	with $\mathbf{A}^\mathbf{k}\equiv(\langle A_0^\mathbf{k}\rangle,\langle A_1^\mathbf{k}\rangle,\langle A_2^\mathbf{k}\rangle)^\text{T}$. We note that eq.~\eqref{matrix_eq} reduces to the conventional equation of motion for the SU(1,1) generators in the $a_i\rightarrow0$ limit~\cite{cheng2021many,lv2020s}.

	The matrix eq.~\eqref{matrix_eq} needs to be solved numerically for the density $n$ is time-dependent. While a lot of information can be extracted by considering the short-time dynamics near an arbitrary time $t_0$ where we may approximate the density by a constant $n(t)\simeq n(t_0)+O(t-t_0)$. In this case, the solution to matrix equation can be written as
	\begin{align}
		\mathbf{A}^\mathbf{k}(t)\simeq\mathbf{A}_s^\mathbf{k}+\sum_{j=0}^2\mathbf{C}_je^{-2i(t-t_0)\xi_{j,\mathbf{k}}}.
	\end{align}

	Here $\mathbf{A}_s^\mathbf{k}$ is the quasi-steady value for the SU(1,1) generators whose elements are listed in the supplementary material, $\mathbf{C}_j$ are constant vectors which depend on the initial value of $\mathbf{A}^\mathbf{k}$ at $t=t_0$, and $2i\xi_{j,\mathbf{k}}$ represent the three eigenvalues of the 3-by-3 matrix in eq.~\eqref{matrix_eq}. The eigenvalues can be calculated explicitly,
	\begin{align}
		\xi_{0,\mathbf{k}}=2i\gamma_0n,\quad\xi_{(1,2),\mathbf{k}}=2i\gamma_0n\pm\sqrt{\epsilon_\mathbf{k}+2g_0n\epsilon_\mathbf{k}-\gamma_0^2n^2}.\nonumber
	\end{align}
	Clearly, $\xi_{(1,2),\mathbf{k}}$ reduce to the excitation energies of Bogoliubov modes in the $\gamma_0\rightarrow0$ limit and the imaginary $\xi_{0,\mathbf{k}}$ indicates that the system only has one true steady state, {\it i.e.} the vacuum~\cite{ness}.

	Even though the whole system eventually evolves to the vacuum state, it is still possible to discuss the stability of the system in short time period $\gamma_0n(0)t\lesssim1$ where many bosons still remain in the system. The excitation energies $\xi_{(1,2),\mathbf{k}}$ provide these stability information.

	Note that a negative imaginary part in $\xi_{j,\mathbf{k}}$ represents an exponentially grow of that mode. In the conventional analysis on BECs no loss ($\gamma_0=0$), the atomic cloud is unstable whenever the argument under the square root is negative for some $\mathbf{k}$, {\it i.e.} when $g_0<0$. However, in the presence of losses ($\gamma_0>0$), there is a competition between the leading $2i\gamma_0n$ term and the imaginary part from the square roots in $\xi_{(1,2),\mathbf{k}}$. This depicts the competition between the particle decay process which stabilizes the system and the collapse process which destabilizes it. For $g_0>-\sqrt{3}\gamma_0$, $\text{Im}(\xi_{(1,2),\mathbf{k}})>0$ for all momenta, the number of excitations always decay and the system keeps evolving towards the quasi-steady state $\mathbf{A}_s^\mathbf{k}$. While for $g_0<-\sqrt{3}\gamma_0$, $\text{Im}(\xi_{(1,2),\mathbf{k}})<0$ for small momenta. The system is unstable against the strong attraction in this region and the Bogoliubov modes as well as the depletion $\frac{1}{N}\sum_\mathbf{k}N_\mathbf{k}\equiv\sum_\mathbf{k}(\langle A_0^\mathbf{k}\rangle-\frac{1}{2})$ keeps growing until the atomic cloud collapses.

	\begin{figure}[t]
	\includegraphics[width=1\linewidth]{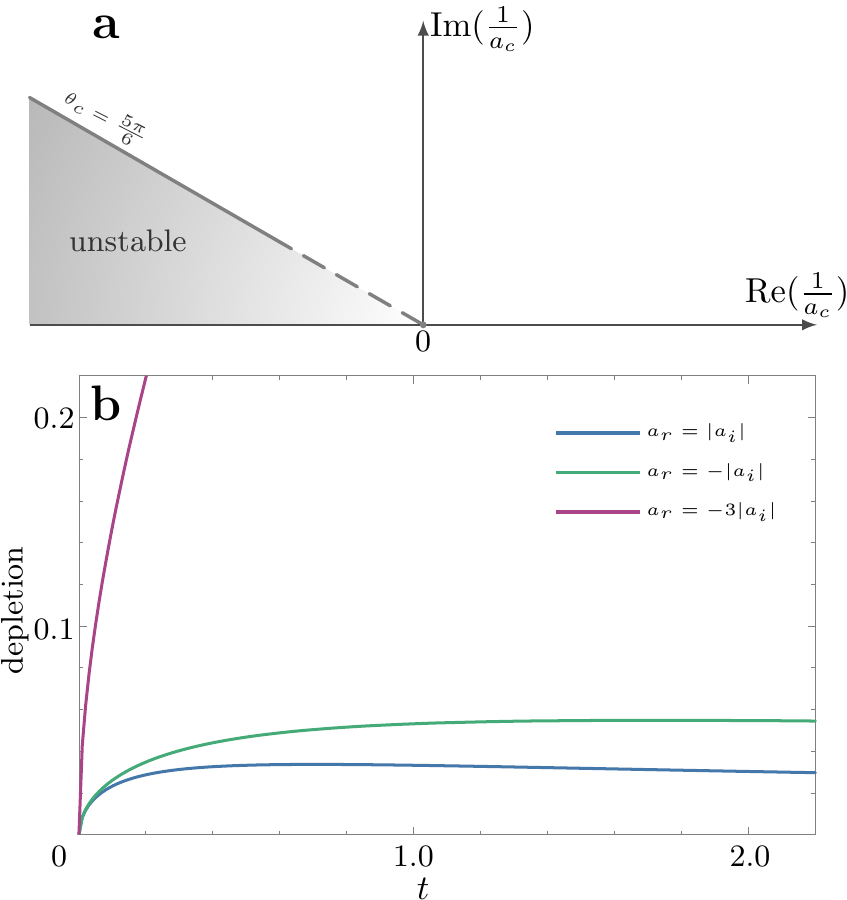}
	\caption{{\bf a}. The phase diagram on the complex $a_c^{-1}$ plane for BECs subjected to weak interaction and two-body loss. The system is unstable for $\theta\equiv\text{arg}(a_c^{-1})>\frac{5\pi}{6}$. {\bf b}. The depletion $\frac{1}{N}\sum_{\mathbf{k}\neq0}N_\mathbf{k}$ as a function of time (in unit of $\frac{1}{\gamma_0n(0)}$). The initial condition is $\mathbf{A}^\mathbf{k}_0=(\frac{1}{2},0,0)^\text{T}$; the parameters are $g_0=\gamma_0$ (blue, stable), $g_0=-\gamma_0$ (green, stable), and $g_0=-3\gamma_0$ (purple, unstable). \label{fig}}
	\end{figure}

	The different behaviors define a critical angle $\theta_c=\frac{5\pi}{6}$ for $\text{arg}(a_c^{-1})$ which separates the complex $a_c^{-1}$ plane into two regions. To demonstrate the difference of dynamics in these regions, we numerically solve the matrix eq.~\eqref{matrix_eq} for different $g_0/\gamma_0$ and plot the depletion as a function of time in Fig.~\ref{fig}b. One can see that for $g_0/\gamma_0<-\sqrt{3}$, the depletion quickly grows and reaches $O(1)$ where the Bogoliubov approximation becomes invalid, in contrast to the cases with $g_0/\gamma>-\sqrt{3}$ where the depletion remains small.

	The total particle number decay rate $\dot{N}$ are also governed by the SU(1,1) generators. It can be shown that $\dot{N}=-2\gamma nN-2\gamma_0n\sum_{\mathbf{k}\neq0}(A_0^\mathbf{k}+A_1^\mathbf{k}-\frac{1}{2})$~\cite{sm}. Note that the expectation value of $\dot{N}$ depends on the specific initial state of the system except the leading mean-field decay $-\gamma nN$. Nevertheless, we may calculate $\dot{N}$ for the quasi-steady state, which helps demonstrate the renormalization relation~\eqref{complex_renormalization}.

	 For the quasi-steady state, we have
	\begin{align}
		\langle\dot{N}\rangle_s=-2\gamma nN+2\gamma_0n\sum_{\mathbf{k}\neq0}\frac{g_0n\epsilon_\mathbf{k}+\gamma_0^2n^2}{\epsilon_\mathbf{k}^2+2g_0n\epsilon_\mathbf{k}+3\gamma_0^2n^2}.
	\end{align}
	Note that the momentum summation on the R.H.S. has a ultraviolet divergence because the leading terms in the summand are of order $1/k^2$ for large $k$. It has the same origin as the divergence appears in the ground state energy in BECs without losses~\cite{pethick2008bose}, and similarly, can be cured using the renormalization relation~\eqref{complex_renormalization} which substitutes the bare coupling constant $\gamma$ by its second order expansion $\gamma_0+\frac{g_0\gamma_0}{\Omega}\sum_\mathbf{k}\frac{1}{\epsilon_\mathbf{k}}$. Finally, we obtain the particle rate to the order of $(n|a_c|^{3})^{1/2}$
	\begin{align}
		\langle\dot{N}\rangle_s=-\frac{8\pi\hbar^2|a_i|nN}{m}\left[1+2\sqrt{2\pi}c_\theta(n|a_c|^3)^{1/2}\right]
	\end{align}
	with $c_\theta=\frac{\cos(2\theta)}{\sqrt{\cos(\theta-\pi/3)}}+2\cos\theta\sqrt{\cos(\theta-\pi/3)}$ and $\theta=\text{arg}(a_c^{-1})\in(0,\theta_c)$.

	We note that the leading term in $\dot{N}$ may be viewed as the mean-field effect due to the two-body loss, which gives particle decay on the mean-field level $n(t)\simeq n(0)/(1+\gamma_0n(0)t)$~\cite{semi_classical}. While the next term in the order of $(n|a_c|^3)^{1/2}$ is an analog to the celebrated Lee-Huang-Yang correction for weakly interacting Bose gas~\cite{lee1957many,lee1957eigenvalues}.

	{\it Outlooks and final remarks.-} 
	Besides its experimental relevance to open systems, the complex contact interaction might also profoundly improve our understanding on close systems. We believe that $\text{Im}(a_c)$ or $\text{Im}(a_c^{-1})$ bear much more deep physical meaning than just a real parameter added to the Hermitian Hamiltonian. This is because it allows the analytical continuation of many physical quantities to the entire complex plane of $a_c^{-1}$, which in turn could help understand the physics on the real axis through their analytic properties. As a simple example, the stability analysis on the complex $a_c^{-1}$ plane (see Fig.~\ref{fig}a) provides a natural explanation of why a regular BEC (without loss) is unstable in the attractive regime ($g_0<0$). Other examples include few-body physics such as the three-body Efimov states in complex plane~\cite{CLS}.

	Finally, we comment on the experiment control of $a_c$. Complex scattering lengths have been observed in cold atom experiments through optical Feshbach resonance~\cite{chin2010feshbach}. As the optical Feshbach resonance couples the open scattering channel to a closed channel molecule with finite lifetime, which results a complex scattering length that can be tuned via controlling the detuning and the intensity of the optical fields. Indeed, we develop a resonant two-channel model with finite lifetime closed channel dimer and show that the complex scattering length $a_c$ can be experimentally tuned across the entire lower half complex plane~\cite{sm}.

	{\it Acknowledgement.-} We wish to express appreciation to Hui Zhai without whom the work would not have been possible. We are also grateful to Haibin Wu, Ran Qi, Zhenhua Yu, Xiaoling Cui, Peng Zhang and Ren Zhang for fruitful discussions.  This work is supported by NSFC under Grant No. 12004115, Program of Shanghai Sailing Program Grant No. 20YF1411600.

	\bibliography{references}

\begin{thebibliography}{30}
\bibitem [{bos()}]{bosons}%
  \BibitemOpen
  \href@noop {} {}\bibinfo {note} {For simplicity, we focus on systems consist
  of identical bosons in this work. Generalizations to fermionic or mixed
  systems are straightforward.}\BibitemShut {Stop}%
\bibitem [{ai()}]{ai}%
  \BibitemOpen
  \href@noop {} {}\bibinfo {note} {Consider integral
  $\int_0^{r_0}r^2drV_i|\varphi|^2=\int_0^{r_0}drV_i|u|^2$ with
  $\varphi(r)\equiv{u(r)}/{r}$. Note that $u(r)$ satisfies the zero-energy
  Schr\"{o}dinger equaiton $-\frac{1}{m}u''+V_c u=0$. We thus have
  $\int_0^{r_0}V_c|u|^2=\frac{1}{m}u^*u'|_{r_0}-\frac{1}{m}\int_0^{r_0}dr|u'|^2$.
  Take the imaginary part and use the fact that $u(r)=1-r/a_c$ for $r\geq r_0$,
  we reach the conclusion
  $\text{Im}(a_c^{-1})=m\int_0^{r_0}V_i|u|^2$}\BibitemShut {NoStop}%
\bibitem [{dro()}]{drop}%
  \BibitemOpen
  \href@noop {} {}\bibinfo {note} {Note that the relative kinetic energy
  diverges in the order of $r_{\alpha\beta}^{-2}$ in this region.}\BibitemShut
  {Stop}%
\bibitem [{nes()}]{ness}%
  \BibitemOpen
  \href@noop {} {}\bibinfo {note} {In comparison, note that a closed Hermitian
  system has more than one steady state such as
  $\hat{\rho}=|\Psi\rangle\langle\Psi|$, where $|\Psi\rangle$ could be an
  arbitrary eigenstate to the Hermitian Hamiltonian.}\BibitemShut {Stop}%
\bibitem [{sem()}]{semi_classical}%
  \BibitemOpen
  \href@noop {} {}\bibinfo {note} {The mean-field decay may also be obtained
  through a semi-classical analysis. First, it can be estimated that the number
  of inelastic collision happened in $dt$ being
  $dN_{\text{collision}}=nv\sigma_{\text{inelastic}}dt$ per particle. Here
  $\sigma_\text{inelastic}=\frac{4\pi}{k}\frac{|a_i|}{(1-ka_i)^2+k^2a_r^2}$ is
  the two-body inelastic scattering cross section and $v$ is the mean relative
  velocity between particles. For each inelastic collision the system loses two
  particles, we thus have
  $\dot{N}/N=-2\lim_{k\rightarrow0}\frac{dN_{\text{collision}}}{dt}=-8\pi\hbar^2|a_i|n/m$.}\BibitemShut
  {Stop}%
\end{thebibliography}

	\begin{widetext}
	\section*{Elements of $\mathbf{A}^\mathbf{k}_s$}
$\mathbf{A}^\mathbf{k}_s$ in the main text is given by
\begin{align}
	\mathbf{A}_\text{s}^\mathbf{k}=\left(\frac{1}{2}+\frac{(g_0^2+\gamma_0^2)n^2}{2\epsilon_\mathbf{k}^2+4g_0n\epsilon_\mathbf{k}+6\gamma_0^2n^2},
	-\frac{g_0n\epsilon_\mathbf{k}+g_0^2n^2+2\gamma_0^2n^2}{2\epsilon_\mathbf{k}^2+4g_0n\epsilon_\mathbf{k}+6\gamma_0^2n^2},
	-\frac{\gamma_0 n(\epsilon_\mathbf{k}-g_0n)}{2\epsilon_\mathbf{k}^2+4g_0n\epsilon_\mathbf{k}+6\gamma_0^2n^2}\right)^\text{T}.\label{steady_value}
\end{align}

\section*{Derivation of the particle rate}

The particle decay rate $\dot{N}$ can be decomposed to the decay rate of condensate particles $N_0$ and the quantum depletion $\sum_{\mathbf{k}\neq0}N_\mathbf{k}$, we thus have
\begin{align}
	\dot{N}=\frac{dN_0}{dt}+\sum_{\mathbf{k}\neq0}\frac{dN_\mathbf{k}}{dt}=\frac{dN_0}{dt}+\sum_{\mathbf{k}\neq0}\frac{dA^\mathbf{k}_0}{dt}.
\end{align}
Here $\frac{dA^\mathbf{k}_0}{dt}$ is already known from the matrix equation in the main text.

To derive the decay rate of condensate particles, we note that
\begin{align}
	\frac{d\langle{N}_0\rangle}{dt}=\partial_t\text{tr}\left(\hat{\rho}\hat{a}_0^\dagger \hat{a}_0\right)=\text{tr}\left(\mathcal{L}(\hat{\rho})\hat{a}^\dagger_0\hat{a}_0\right)=\text{tr}\left(\hat{\rho}\mathcal{L}'(\hat{a}_0^\dagger \hat{a}_0)\right),
\end{align}
with $\mathcal{L}'$ defined by
\begin{align}
	\mathcal{L}'(\hat{O})\equiv i\left[\hat{H},\hat{O}\right]-\frac{\gamma}{2\Omega}\sum_{\mathbf{k,k',p}}\left\{\hat{a}^\dagger_{\mathbf{k+p}}\hat{a}^\dagger_{\mathbf{k'-p}}\hat{a}_\mathbf{k'}\hat{a}_\mathbf{k},\hat{O}\right\}+\frac{\gamma}{2V}\sum_{\mathbf{k,k',p}}\hat{a}^\dagger_{\mathbf{k+p}}\hat{a}^\dagger_{\mathbf{k'-p}}\hat{O}\hat{a}_\mathbf{k'}\hat{a}_\mathbf{k}.
\end{align}
In fact the above equation may be regarded as the complex analog of the Heisenberg equation for an open system. 

Within the Bogoliubov approximation, it can be shown that $\mathcal{L}'(\hat{a}_0^\dagger\hat{a}_0)$ can also be expressed by the SU(1,1) generators,
\begin{align}
	\mathcal{L}'(\hat{a}^\dagger_0 \hat{a}_0)\simeq -2\gamma nN-2\sum_{\mathbf{k}\neq0}\left(\gamma_0nA_1^\mathbf{k}+g_0nA_2^\mathbf{k}\right).
\end{align}

Together with the expression for $\frac{dA_0^\mathbf{k}}{dt}$, we have
\begin{align}
	\dot{N}=-2\gamma nN +2\gamma_0n\sum_{\mathbf{k}\neq0}\left(A_0^\mathbf{k}+A_1^\mathbf{k}-\frac{1}{2}\right).
\end{align}
We note that the R.H.S. is equivalent to two times the imaginary part of the effective Hamiltonian $H_\text{eff}$ under Bogoliubov approximation.

\section*{Two-channel model realization of complex $a_c$ }
Now if we introduce a bosonic $\hat{d}$ field to describe the molecule in the closed channel, the two-channel Hamiltonian can be written as 
\begin{equation}
\hat{H}_{\text{two-channel}} = \sum_{\mathbf{k}}\left(\epsilon_{\mathbf{k}} \hat{b}_{\mathbf{k}}^{\dagger}\hat{b}_{\mathbf{k}} +\xi_{\mathbf{k}}\hat{d}_{\mathbf{k}}^{\dagger}\hat{d}_{\mathbf{k}}\right)+ \frac{g}{2\Omega}\sum_{\mathbf{k},\mathbf{k'},\mathbf{p}}\hat{b}^{\dagger}_{\mathbf{k}+\mathbf{p}}\hat{b}^{\dagger}_{\mathbf{k'}-\mathbf{p}}\hat{b}_{\mathbf{k'}}\hat{b}_{\mathbf{k}}  + \frac{1}{\sqrt{\Omega}}\sum_{\mathbf{k},\mathbf{p}}\left(\alpha \hat{d}_{\mathbf{p}}^{\dagger}\hat{b}_{\frac{\mathbf{p}}{2}-\mathbf{k}}\hat{b}_{\frac{\mathbf{p}}{2}+\mathbf{k}}+\alpha^{*} \hat{d}_{\mathbf{p}}\hat{b}_{\frac{\mathbf{p}}{2}-\mathbf{k}}^{\dagger}\hat{b}_{\frac{\mathbf{p}}{2}+\mathbf{k}}^{\dagger}\right) 
\end{equation}
with $\epsilon_{\mathbf{k}} = \frac{k^2}{2m}$ and $\xi_{\mathbf{k}} = \frac{k^2}{4m}+\nu$.

To put in a two-body loss term, we consider following Lindblad master equation,
\begin{equation}
\partial_{t}\hat{\rho} = \mathcal{L}(\hat{\rho}) = \frac{1}{i}[ \hat{H}_{\text{two-channel}},\hat{\rho}]-\frac{\gamma}{2}\sum_{\mathbf{k}}\{\hat{d}^{\dagger}_{\mathbf{k}}\hat{d}_{\mathbf{k}},\hat{\rho}\} + \gamma\sum_{\mathbf{k}}\hat{d}_{\mathbf{k}}\hat{\rho} \hat{d}_{\mathbf{k}}^{\dagger}.
\end{equation}

Using the argument in the main text, we know that the dynamics of the two-body density matrix is equivalent to the evolution under a non-hermitian Hamiltonian
\begin{equation}
\hat{H}_\text{eff} = \hat{H}_{\text{two-channel}} - i\frac{\gamma}{2}\sum_{\mathbf{k}} \hat{d}_{\mathbf{k}}^{\dagger}\hat{d}_{\mathbf{k}}.
\end{equation}

We can obtain the two-body scattering matrix $T_{2}$ for this $\hat{H}_\text{eff}$
\begin{equation}
T_{2}(E) = \left(\left(g +\frac{|\alpha|^2}{E-\nu+i\frac{\gamma}{2}}\right)^{-1}-\frac{1}{\Omega} \sum_{\mathbf{k}}\frac{1}{E-\hbar^2 k^2/m}\right)^{-1}.
\end{equation}
Here the complex $a_{c}(E)$ depends on $E$ and is related to $T_{2}$ as $T_{2}(E) = \frac{4\pi \hbar^2}{m}\left(\frac{1}{a_{c}(E)} + i\sqrt{mE/\hbar^2}\right)^{-1}.$

The renormalize relation is then given by
\begin{equation}
\frac{m}{4\pi\hbar^2 a_{c}(E)} =\left(g+\frac{|\alpha|^2}{E-\nu+i\frac{\gamma}{2}}\right)^{-1} + \frac{1}{\Omega} \sum_{\mathbf{k}}\frac{m}{\hbar^2 k^2},
\end{equation} 
where $a_{c}(E)$ can be further written as 
\begin{equation}
a_c(E)=a_\text{bg}+\frac{m}{4\pi\hbar^2}\frac{|\alpha_{\text{re}}|^2}{E-\nu_{\text{re}}+i\frac{\gamma}{2}}
\end{equation}
with $\frac{m}{4\pi \hbar^2 a_{\text{bg}}} = \frac{1}{g} + \frac{m\Lambda}{2\pi^2\hbar^2}$, $
|\alpha_{\text{re}}|^2 = \frac{|\alpha|^2}{(1+mg\Lambda/(2\pi^2\hbar^2))^2}$, 
 $\nu_{\text{re}} = \nu - \frac{m\Lambda |\alpha|^2}{2\pi^2\hbar^2 + mg\Lambda}$ ($\Lambda$ is the momentum cut-off).

Taking $\Lambda \to \infty$ and defining $\Gamma(I) = \frac{m|\alpha_{\text{re}}|^2}{4\pi \hbar^2 a_{\text{bg}}}$, we have 
\begin{equation}
a_c(E) = a_{\text{bg}}\left(1 + \frac{\Gamma(I)}{E- \nu - \Gamma(I)+i\frac{\gamma}{2}}\right).~\label{final}
\end{equation}

We have thus obtained the same result as in Ref.~\cite{chin2010feshbach} which is derived from a multi-channel finite-range model. $\Gamma(I)$ is linear in the laser density $I$ since it is proportional to square of the renormalized coupling strength $\alpha_\text{re}$, and $\nu$ represents the unshifted detuning between the molecule state and the collisional state of the two atoms at $E=0$. Based on eq.~\eqref{final}, it can be shown that $a_c^{-1}(E=0)$ can be tuned across the entire upper half complex plane via controlling the intensity $I$ and the detuning $\nu$ of the laser field.

	\end{widetext}
\end{document}